\documentclass[cits]{PoS}
\ShortTitle{The Murchison Widefield Array}

\title{The Murchison Widefield Array}

\author{\speaker{Daniel A. Mitchell}\thanks{This work uses data obtained from the Murchison Radio-astronomy Observatory.
        We acknowledge the Wajarri Yamatji people as the traditional owners of the Observatory site.  Support came from
        the U.S. National Science Foundation (grants AST-0457585 and PHY-0835713), the Australian Research Council
        (grants LE0775621 and LE0882938), the U.S. Air Force Office of Scientific Research (grant FA9550-0510247), the
        Smithsonian Astrophysical Observatory, the MIT School of Science, the Raman Research Institute, the Australian
        National University, the iVEC Petabyte Data Store, the Initiative in Innovative Computing and NVIDIA sponsored
        Center for Excellence at Harvard, and the International Centre for Radio Astronomy Research, a Joint Venture of
        Curtin University of Technology and The University of Western Australia, funded by the Western Australian State
        government. The speaker would also like to thank the National Astronomy and Ionosphere Center and Cornell
        University for travel support under grant AST-1013766.}, Lincoln J. Greenhill, Stephen M. Ord and Gianni
        Bernardi\\

        Harvard-Smithsonian Center for Astrophysics,
        e-mail: \email{dmitchell@cfa.harvard.edu}}

\author{Randall B. Wayth\\
        Curtin University of Technology}

\author{Richard G. Edgar, Michael A. Clark, Kevin Dale and Hanspeter Pfister\\
        Harvard University}

\author{Stewart J. Gleadow\\
        University of Melbourne}

\author{W.~Arcus, F.H.~Briggs, L.~Benkevitch, J.D.~Bowman, J.D.~Bunton, S.~Burns, R.J.~Cappallo, B.E.~Corey,
        A.~de~Oliveira-Costa, L.~Desouza, S.S.~Doeleman, M.F.~Derome, D.~Emrich, M.~Glossop, R.~Goeke , M.R.~Gopala
        Krishna, B.~Hazelton, D.E.~Herne, J.N.~Hewitt, P.A.~Kamini, D.L.~Kaplan, J.C.~Kasper, B.B.~Kincaid, J.~Kocz,
        E.~Kowald, E.~Kratzenberg, D.~Kumar, C.J.~Lonsdale, M.J.~Lynch, S.~Madhavi, M.~Matejek, S.R.~McWhirter,
        M.F.~Morales, E.~Morgan, D.~Oberoi, J.~Pathikulangara, T.~Prabu, A.~Rogers, J.E.~Salah, R.J.~Sault,
        N.~Udaya~Shankar, K.S.~Srivani, J.~Stevens, S.J. Tingay, A.~Vaccarella, M.~Waterson, R.L.~Webster, A.R.~Whitney,
        A.~Williams and C.~Williams\\ www.mwatelescope.org}

\abstract{It is shown that the excellent Murchison Radio-astronomy Observatory site allows the Murchison Widefield Array
          to employ a simple RFI blanking scheme and still calibrate visibilities and form images in the FM radio band.
          The techniques described are running autonomously in our calibration and imaging software, which is currently
          being used to process an FM-band survey of the entire southern sky.}

\FullConference{RFI mitigation workshop\\
                29-31 March 2010\\
                Groningen, the Netherlands}

\begin{document}

\section{Introduction}

The Murchison Widefield Array (MWA) is both a pathfinder to the Square Kilometre Array and a stand-alone instrument,
designed for cutting-edge science between 80 and 300 MHz via real-time imaging. This frequency band is rich with
scientific potential, key science drivers being detection of redshifted 21 cm emission from neutral Hydrogen during the
Epoch of Reionization, detection of radio transients, and space weather monitoring. However, the band can be difficult
to observe in due to high levels of RFI. To reduce the need for complicated interference mitigation strategies, a 
remote site has been chosen so that RFI-free parts of the spectrum are available across the entire frequency range.
While the 1024-input correlator and high-throughput software backend make the MWA a state-of-the-art instrument for
advanced cancellation techniques, simple blanking methods have the advantage of simple statistics and are well suited to
the low spectral and temporal occupancy at the site. Maintaining simple statistics is particularly important in the FM
radio band, where it is anticipated that part of the weak neutral Hydrogen signal emitted during the Epoch of
Reionization lies.

In addition to the excellent site, several design features help the MWA confront other challenges faced by modern and
future radio facilities. For example, restricting the maximum baseline to a few kilometres reduces ionospheric
distortion models to two dimensions, and real-time snapshot imaging allows the removal of these and various wide-field
distortions in the image plane. Real-time processing also enables the large visibility data rate that comes with 512
antennas. A consequence of real-time processing and high data rates is that RFI algorithms need to be robust and
efficient. A description of one such approach follows.

\subsection*{The Murchison Radio-astronomy Observatory}

The Murchison Radio-astronomy Observatory (MRO) is a protected site in remote Western Australia that has been set aside
as a park for a number of next-generation radio facilities. It consists of a zone, with a radius of 100 to 150 km, in
which radio emission is restricted. While this level of protection is imperative for detection of extremely weak target
signals, terrestrial radio waves can travel large distances via reflections off aircraft or meteor trails, or by
occasional refractive ducting, while signals from satellites and aircraft can be in the direct line of sight.

FM radio signals from communities outside the restricted zone are an example of signals that will at times be reflected
into the MRO. However, removing a modest amount of contaminated data will preserve this important band. A neighbouring
instrument, the Experiment to Detect the Global Epoch of Reionization Signature (EDGES), removed all detected FM
interference by filtering 3 percent of their 130 MHz band for up to 15 percent of the time during a typical day, with
statistics collected over approximately 80 days \cite{Rogers2010}.

\section{The Real-Time Calibration and Imaging System}

The MWA will consist of 512 antenna tiles, each a 4x4 array of dual-polarised, vertical bowtie dipoles. It is currently
in a final design phase, with the first 32 antennas -- along with prototype beamformers, receivers, correlator boards
and computing hardware -- being regularly used in an end-to-end pipeline. The primary aim of the 32-tile system is to
test hardware and firmware before the build out of the full array, and to aid the design of real-time software systems.

With the challenging dynamic range goal of a million to one in mind, a comprehensive, wide-field, real-time calibration
and imaging system has been designed (see \cite{Mitchell2008} and \cite{Lonsdale2009}). While the bulk of the real-time
system is running routinely on simulated data using clusters with substantial graphics-card acceleration \cite{Edgar},
for the 32-tile array the design has been significantly simplified. In keeping with the series of papers that accompany
\cite{Hamaker1996}, and if the visibilities are dominated by a strong, unresolved, calibrator source, one can write the
measured coherency matrix (containing the four polarisation terms of the visibility), after passing through antennas $j$
and $k$, as

\begin{equation}
\begin{array}{lcl}
{\bf V}_{jk} = {\bf J}_{j}\, \left( {\bf S}\exp\{-i\phi_{jk}\} \right) \,{\bf J}_{k}^\dagger + {\bf N}_{jk},
\end{array}
\label{visibility}
\end{equation}

\noindent where a dagger superscript denotes a conjugate transpose, $\phi_{jk}$ is the phase shift due to differences in
the path length from the source to the two antennas, and each bold symbol in (\ref{visibility}) represents a $2\times2$
matrix; ${\bf S}$ is the coherency matrix of the calibrator signal, the Jones matrices, ${\bf J}_{j}$ and ${\bf J}_{k}$,
describe the response of the antennas, and ${\bf N}_{jk}$ contains the covariance of the polarised components of the
noise. This noise comes mainly from the receivers and from other radio sources and, in the absence of RFI and
significant secondary sources, we might expect it to have a Gaussian distribution. This is discussed further in section
\ref{results}.

With a model for the source, ${\bf S}\exp\{-i\phi_{jk}\}$, and noise with an approximately Gaussian distribution, we can
estimate the Jones matrix for each of the $M$ antennas towards the calibrator using a least-squares approach (any
amplitude or phase errors in the source model will be absorbed into these estimates). Following the matrix formalism
suggested in \cite{Hamaker2000} and described in \cite{Mitchell2008}, we search for the matrices $\widehat{{\bf J}}_{j}$
that minimise

\begin{equation}
\chi^2 = \sum_{j=1}^{M}\;\sum_{k,\,k\neq j}^{M}\sigma_{jk}^{-2}
\left\|{\bf V}_{jk}\,\exp\{+i\phi_{jk}\} -
       \widehat{{\bf J}}_{j}\,{\bf S}\,\widehat{{\bf J}}_{k}^\dagger\right\|_F^2,
\label{chiSq}
\end{equation}

\noindent where, for arbitrary matrix ${\bf A}$, the squared Frobenius norm $\|{\bf A}\|_F^2=\mbox{Tr}\left({\bf
A}^{}{\bf A}^\dagger\right)$, and $\sigma_{jk}^{2}=\left<\|{\bf N}_{jk}\|_F^2\right>$. The solutions of (\ref{chiSq}),

\begin{equation}
\widehat{{\bf J}}_{j}
 = \left(\sum_{k,\,k\neq j}^{M}
   \sigma_{jk}^{-2}\left({\bf V}_{jk}\,\exp\{+i\phi_{jk}\}\right)\widehat{{\bf J}}_{k}\,{\bf S}^\dagger\right)
   \left(\sum_{k,\,k\neq j}^{M}
   \sigma_{jk}^{-2}{\bf S}\,\widehat{{\bf J}}_{k}^\dagger\,\widehat{{\bf J}}_{k}\,{\bf S}^\dagger \right)^{-1},
\label{matrix-solution}
\end{equation}

\noindent can be estimated from an initial set of Jones matrices and iterated until they converge to an optimal set,
which can then be tracked.

The MWA is a snapshot instrument that uses (\ref{visibility}) and (\ref{matrix-solution}) to produce calibrated images
every eight seconds for each frequency channel and polarisation.\footnote{A separate algorithm is used to estimate
deviations in $\phi_{jk}$ arising from the ionosphere.} The images are regridded in real time to remove ionospheric and
wide-field distortions, then integrated for several minutes. Eventually (\ref{matrix-solution}) will be updated for many
calibrators, to build up direction-dependent gain models, but for the 32-tile array it is assumed that the shape of the
primary beam response of each antenna is known, so that the models can be fixed by a single calibrator. It is also
assumed that the primary beam shapes are the same for each antenna, so that the fixed beams are equal. These issues are
described in more detail in \cite{Ord2010}, where the real-time system was applied to the 32-tile array to track fields
and integrate for many hours over much of the band from 90 to 210 MHz.

\subsection{Flagging in the Real-Time System}

With all of the dimensions that can be used to describe a typical observation -- time, frequency, polarisation, antenna
space, visibility space, image space, etc. -- RFI mitigation will in general be most effective if one can find a
projection in which the interference has little or no overlap with the signals of interest. For RFI that is contained
within small segments of time and frequency, these two dimensions are obvious places to define the segmentation, and
traditional techniques for blanking samples in time and frequency offer close to complete removal of the RFI at the
expense of losing a small amount of the signal of interest.

The calibration and imaging system has a separate processing pipeline for each frequency channel, and this coupled with
the real-time nature of the instrument makes it well suited to time and frequency blanking. While there are a number of
places to set flags that can be used throughout the system to blank visibilities, it is only being done at the moment at
the start of the software system. There are two detectors in use; one generates statistics for each frequency channel
using neighbouring time samples, the other generates statistics for each time step using neighbouring frequency samples.
Both detectors search for outliers using quartiles, where departures from the median value by more than a set multiple
of the interquartile range are flagged for blanking. The interquartile range is a robust estimator of signal variance
and is relatively cheap to compute (see, for example, \cite{Fridman2008}). Each visibility has an associated
non-negative inverse-variance weight, and flags are set by making these negative. The calibration and imaging routines
ignore samples that have a negative weight, and accumulated quantities that are missing samples due to flagging can
simply be down weighted.

Development is underway to combine these detectors with goodness-of-fit estimates for the relations given at the start
of this section. Also, since images are formed every eight seconds, various image processing tools can be run before
integration, as a final check of data quality.

\section{32-Tile Results}
\label{results}

The accuracy of calibration relations (\ref{visibility}) and (\ref{matrix-solution}) will in general be diminished by
any RFI in the coherency matrices. In this section some examples will be given showing the effect of both RFI and
blanking. Data come from 21 of the 32 antennas and 6.2 MHz of bandwidth centred at 98 MHz. The antenna tiles have been
steered to a field with a right ascension of 21.2 hours and a declination of -25.4 degrees, close to transit. The
dominant object in the field is extragalactic radio source PKS J2107-2526, which is 61 Jy at 80 MHz \cite{pkscat90}.
This is one of the calibration pointings from a survey of the entire southern sky, which is currently under
construction.

\begin{figure}
  \parbox[b]{.43\linewidth}{
          \includegraphics[width=\linewidth]{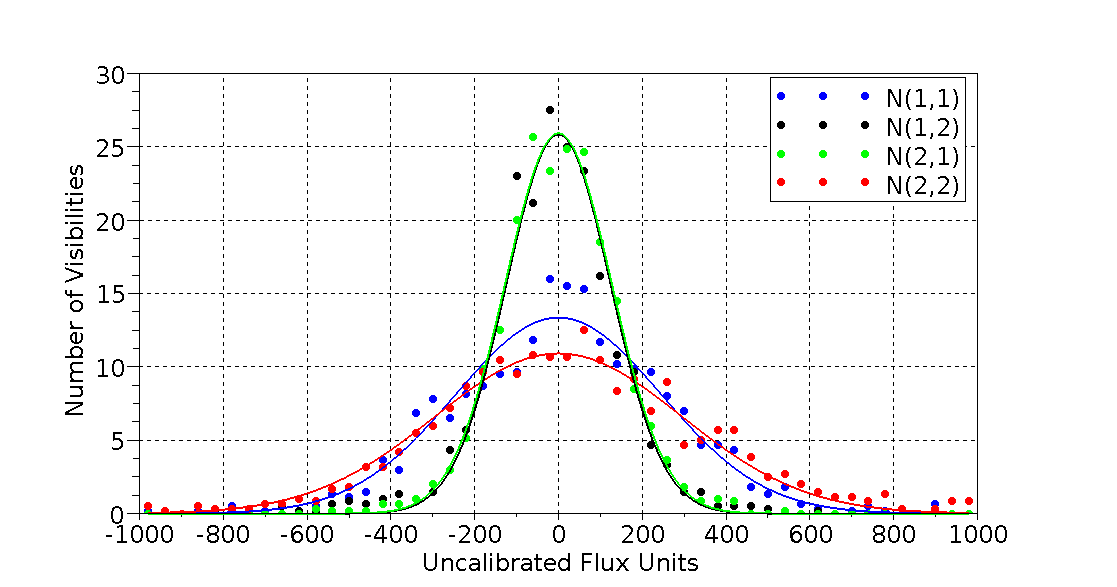}}\hfill
  \parbox[b]{.57\linewidth}{

  \caption{Histograms of the noise in each element of the coherency matrices -- estimated by peeling the fitted source
           contribution -- along with best-fit Gaussian curves. The peeling was done separately in each 40 kHz channel,
           and the statistics generated after combining the whole band (after dividing out a polynomial bandpass fit).
           Several histograms were averaged to reduce the scatter.}\label{noise}}

\end{figure}

In the absence of RFI, the validity of the model given in (\ref{visibility}) may be verified by inspecting noise
matrices. The matrices that minimise (\ref{chiSq}) can be used to remove the calibrator power from the measured
coherency matrix (known as peeling \cite{Noordam2004}), and if the assumptions hold then the remaining noise term should
have a Gaussian distribution. The noise distribution after peeling is shown in figure \ref{noise}. In this paper the
fundamental accuracy of the calibration process is not being considered, and it is adequate to note that the noise
distributions qualitatively match Gaussians.

\begin{figure}[h]
  \centering
  \includegraphics[width=0.43\textwidth]{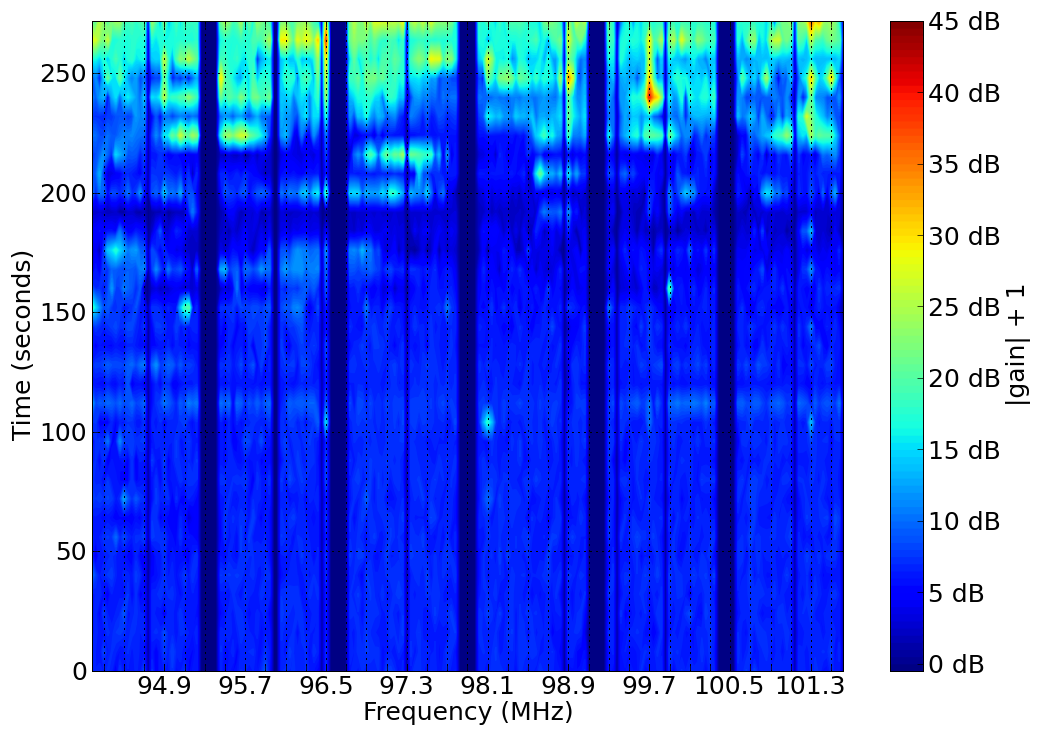}
  \includegraphics[width=0.43\textwidth]{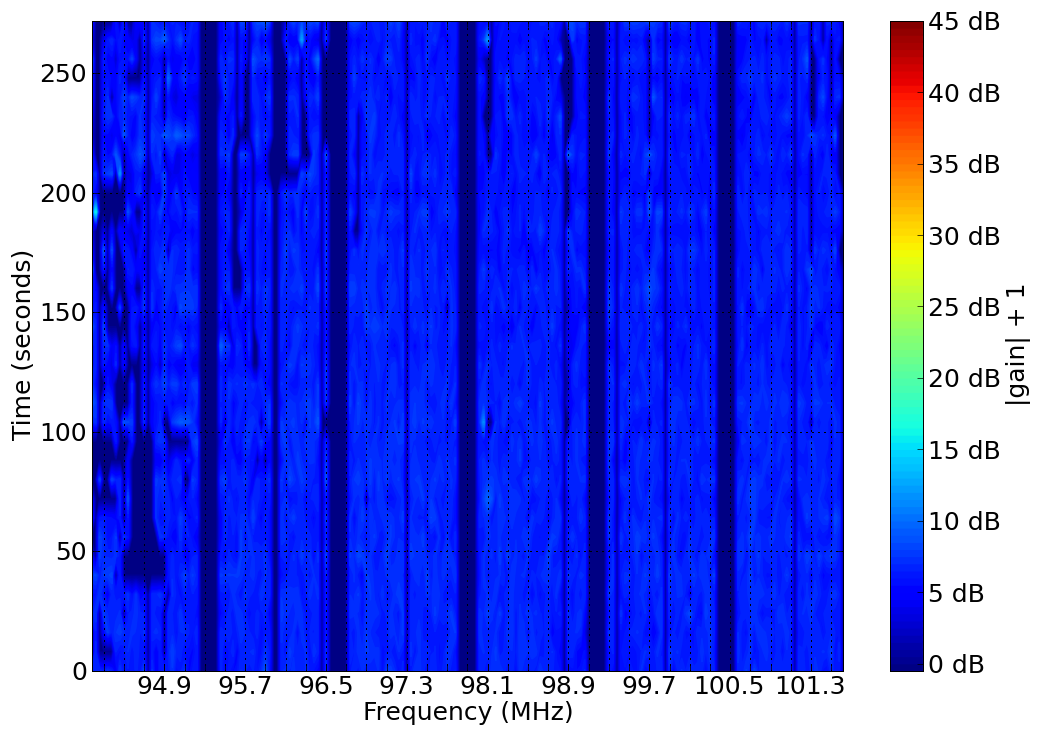}

  \caption{Jones matrix element $J_{j}(1,1)$ is shown as a function of time and frequency for MWA antenna tile 1, with
           dashed vertical lines indicating possible FM radio frequencies. The image on the left-hand side shows fitted
           gain amplitude without RFI blanking, while the image on the right-hand side shows the gain amplitude after
           all detected outliers were blanked. Blanking has maintained the smooth gain structure into the time interval
           that contains the strong RFI, and the entire time interval can be accurately calibrated and imaged.}

  \label{JonesPX}
\end{figure}

In figure \ref{JonesPX} the amplitude of Jones matrix element $J_{j}(1,1)$ is shown as a function of time and frequency
for one of the MWA antennas. To avoid known artifacts of the prototype receiving system, specific frequency channels
were always blanked for these data, which can be seen as vertical bands of zero gain. About halfway though the scan RFI
becomes evident and increases in power as the scan continues (the gain between FM radio frequencies is also affected by
contamination of the bandpass fits).

\begin{figure}
  \centering
  \includegraphics[width=0.33\linewidth]{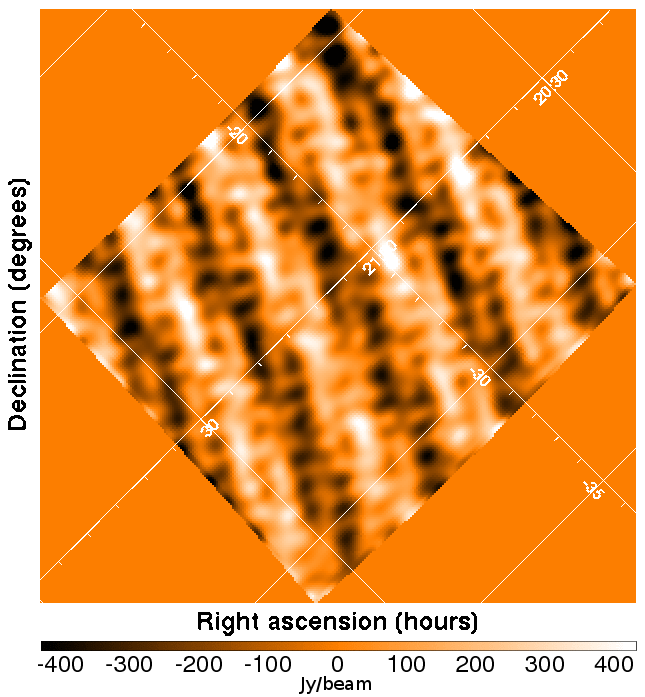}
  \includegraphics[width=0.33\linewidth]{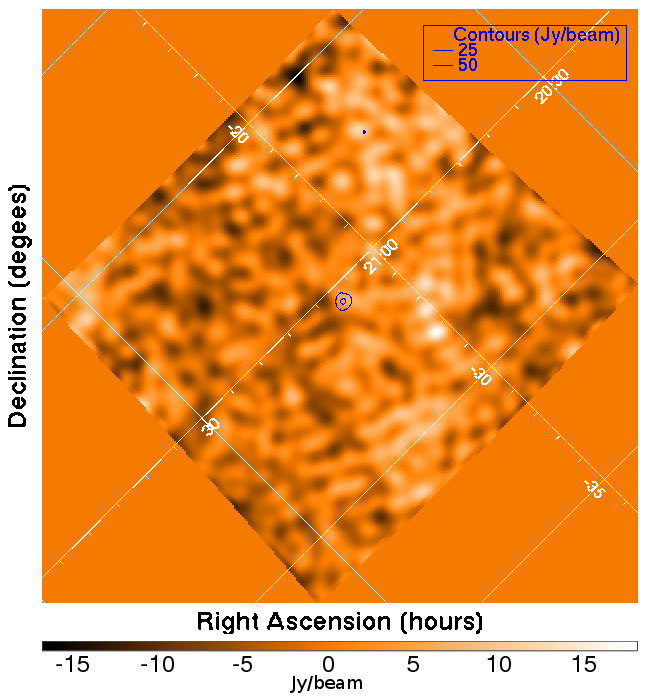}

  \caption{On the left is an image of the J2107-2526 field produced by integrating 8-second snapshots over the entire
           time interval without blanking. On the right is an image of the field after RFI blanking and peeling, along
           with contours of the unpeeled image.}

  \label{image}
\end{figure}

It is clear that blanking has significantly reduced the RFI in these data. Jones matrix errors have been reduced by 30
dB, leaving gains that are stable in both frequency and time. During the worst of the RFI only about 12 percent, or
$\sim$0.76 MHz of bandwidth, was blanked. The quality of the unblanked data can be seen in figure \ref{image}, where
images made with and without blanking are compared. The calibrator is completely buried in the presence of RFI, but can
be imaged and peeled below the noise floor with the RFI flags.

\section{Conclusion}

The MWA employs a real-time calibration and imaging system to cope with a large data rate and fast-changing calibration
parameters. While the MWA is situated in the radio-quiet Murchison Radio-astronomy Observatory, it will observe in
frequency bands that are occupied by numerous communication signals. While these signals are transmitted from distant
locations, they will, on occasion, reflect or refract into the receivers at levels that are orders of magnitude above
the noise floor. During deep integrations the MWA real-time system will simply discard dubious data. This will require a
series of data-quality tests, of which the simple median-based detector shown here will form an integral part.

\end{document}